\documentclass[10pt,pra,twocolumn,superscriptaddress,floatfix,longbibliography]{revtex4-2}

\usepackage{graphicx}
\usepackage{amssymb}
\usepackage{times}
\usepackage{amsmath}
\usepackage{amsthm}
\usepackage{epstopdf}

\usepackage{hyperref}
\usepackage{xcolor}
\input epsf.tex
\usepackage{graphicx}
\usepackage{amsthm}
\usepackage{subfigure}

\begin{document}

\title{Practical security of  local local 
oscillator continuous-variable quantum key distribution systems with pulse width mismatch }

\author{Yi Zheng}\thanks{yizheng@nwpu.edu.cn}
\affiliation
 {School of Computer Science, Northwestern Polytechnical University, Xi'an 710129, Shaanxi, China
}
\author{Jiarui Wu}
\affiliation
 {School of Computer Science, Northwestern Polytechnical University, Xi'an 710129, Shaanxi, China
}
\author{Chenlei Fang}
\affiliation
 {School of Computer Science, Northwestern Polytechnical University, Xi'an 710129, Shaanxi, China
 }
\author{Qingbing Ji}\thanks{jqbxy@163.com}
\affiliation
 {National Key Laboratory of Security Communication, Chengdu 610041, Sichuang, China
 }
\author{Wei Pan}
\affiliation
 {School of Computer Science, Northwestern Polytechnical University, Xi'an 710129, Shaanxi, China
}
\author{Haobin Shi}
\affiliation
 {School of Computer Science, Northwestern Polytechnical University, Xi'an 710129, Shaanxi, China
}

\begin{abstract}
In continuous-variable quantum key distribution (CVQKD) systems, using a local local oscillator (LLO) scheme removes the local oscillator side channel, enhances Bob’s detection performance and reduces the excess noise caused by photon leakage, thereby effectively improves the system’s security and performance. However, in this scheme, since the signal and the LO are generated by different lasers and the signal propagates through the untrusted quantum channel, their pulse widths may become mismatched. 
Such mismatches may reduce the precision of detection at Bob, affecting parameter estimation processes and leading to inaccurate calculations of the secret key rate. Moreover, mismatches may introduce potential security loopholes. Thus, this paper investigates the practical security issues of the LLO-CVQKD system when pulse width mismatch occurs between the local oscillator and the signal. 
We first model the case of pulse width mismatch and analyze its impact on Bob’s detection. Then, we simulate the secret key rate under different mismatch levels.
Based on the analysis, we find that under such mismatch, the key parameters involved in secret key rate calculation are incorrectly estimated, leading to an overestimation of the system’s secret key rate. 
Therefore, this imperfect mismatch can open a loophole for Eve to perform attacks in practical systems. 
To close this loophole, we design a scheme at Bob to monitor the pulse widths of both the signal and the local oscillator, and to reshape the waveform of the local oscillator so that the two lights are matched again. This method eliminates the adverse effects caused by pulse width mismatch and effectively resists Eve’s attacks which exploit this loophole.

\end{abstract}

\maketitle

\section{Introduction}\label{sec1}
In the field of quantum cryptography, quantum key distribution (QKD) is a technology that utilizes the principles of quantum mechanics to achieve secure communication \cite{r1,r2,r3}. This technology has become relatively mature and has been theoretically proven to be absolutely secure \cite{r4,r5,r6,r7}. Currently, QKD systems can be mainly divided into two categories: discrete-variable quantum key distribution (DVQKD) and continuous-variable quantum key distribution (CVQKD). Compared with DVQKD, CVQKD uses continuous-variable quantum states, such as the amplitude and phase of light fields, for key distribution, and CVQKD systems using weak coherent states and a homodyne detector can be well compatible with the existing optical communication systems \cite{r8,r9}. Therefore, it is important to conduct further research on CVQKD.

In CVQKD, schemes based on Gaussian-modulated coherent states are well-known and have already been implemented in many laboratory and field experiments \cite{r10,r11,r12,r13,r14,r15,r16,r17}. In theory, the GMCS method has been proven to be secure and can be used to prevent both collective and coherent attacks \cite{r18,r19}.
However, in practical systems, there are no perfect experimental devices \cite{r20,r21,r22}, and these imperfections in the devices may lead to a degradation on system performance. To enhance the performance of CVQKD, some researchers have introduced non-Gaussian operations, such as photon subtraction and photon addition operations, as well as optical amplification and optical quantum catalysis operations \cite{r23,r24,r25,r26,r27,r28,r29,r30}. These operations have enhanced the system performance. But in standard CVQKD systems, there are not only performance limitations but also practical security issues, such as wavelength attack \cite{r36}, saturation attack \cite{r39} and so on.
Thus, some researchers are also devoted to improving standard CVQKD systems, such as CV-MDI-QKD \cite{r31, r32} and local local oscillator (LLO) CVQKD \cite{r33,r34}. In CVQKD systems, the local oscillator (LO) is an essential component for detection.
In standard CVQKD systems, the LO is generated by Alice and transmitted together with the signal through the insecure channel. However, this transmission method of the local oscillator introduces many problems. Before the advent of local local oscillator CVQKD systems, many attack methods are related to the security issues of the LO \cite{r35,r36,r37}. To address the above issues, the LLO-CVQKD scheme has been proposed. LLO-CVQKD improves the standard CVQKD system by generating the LO locally at Bob, and removes the need for Alice to send a reference LO. 
To align the phase, Alice can send extra reference pulses, and Bob uses them to measure and correct the phase drift between his local oscillator and Alice’s signal. And it can reduce the risk of LO-based attacks and makes the system more secure and practical.

However, in practical LLO-CVQKD systems, there are still challenges. Since the LO and quantum signal are generated by different lasers, there may be discrepancies in their pulse widths. This pulse width mismatch not only affects the detection efficiency at Bob, thereby impacting the secret key rate and transmission distance, but it may also introduce security loopholes, providing Eve with potential opportunities for attacks. There can be many reasons for this mismatch. On one hand, it can be caused by the finite linewidth, frequency drift, and power fluctuations of different lasers; on the other hand, the signal may be influenced by atmospheric and environmental effects during transmission, which may further alter its temporal mode. Moreover, due to the fabrication processes and tuning precision of practical optical components, slight mismatches may also occur when generating or adjusting the LO and signal. When the pulse widths of the signal and the LO are mismatched,  not only does this affect the system performance but also may cause security loopholes. Based on above analysis, in this paper we first establish a theoretical model for pulse width mismatch between the signal and LO in LLO-CVQKD systems. This model illustrates the detection situation when the pulse widths of the signal and the local oscillator at Bob are mismatched. We then provide a mathematical characterization of the impact of this mismatch on the system. And based on the theoretical model, we perform a parameter estimation which quantitatively analyzes the impact of pulse width mismatch on the security of the system. Finally, we conduct simulations to analyze the impact of different mismatch levels on the measurements at Bob. As a result, we find that such pulse width mismatch causes deviations in Bob’s measurement of the quantum signals, and these deviations increase as the degree of mismatch becomes more worse. Besides, the system security is more severely overestimated by pulse width mismatch when the total excess noise is relatively high. Furthermore, we propose a countermeasure to prevent this issue by monitoring the pulse widths of both signal and LO. We monitor and compare the pulse widths of the signal and LO,  whenever a mismatch is detected, we adjust the LO pulse width to realign it with the signal pulse. In this way, this method ensures that no pulse width mismatch occurs at the detection process.

This paper is organized as follows. In Sec.\ref{sec2}, we describe the LLO-CVQKD system and analyze the impact of the mismatch between signal and LO. Then, we do the parameter estimation in Sec.\ref{sec3}. In Sec.\ref{sec4}, We further analyze the impact of pulse width mismatch from the perspective of the secret key rate. And in Sec.\ref{sec5}, we introduce our countermeasure to eliminate this issue. Finally, conclusions are presented in Sec.\ref{sec6}.

\section{LLO-CVQKD with pulse width mismatch }\label{sec2}
In this section, we first introduce the LLO-CVQKD system. Then, we analyze the situation that may arise in a practical LLO-CVQKD system when the pulse widths of the LO and signal mismatch during measurement at Bob.

\subsection{CVQKD with a locally generated local oscillator }
Before introducing the LLO-CVQKD system , we first introduce the standard CVQKD system. The standard CVQKD system is typically based on Gaussian modulation. It is primarily divided into three main components: the transmitter, the channel, and the receiver. The transmitter, Alice, is responsible for preparing the quantum states and encoding the key information. The receiver, Bob, is in charge of measuring the quantum states and decoding the key information. The channel serves as the bridge for transmitting the quantum states from the transmitter to the receiver.

\begin{figure}[!h]\center
\centering
\resizebox{8cm}{!}{
\includegraphics{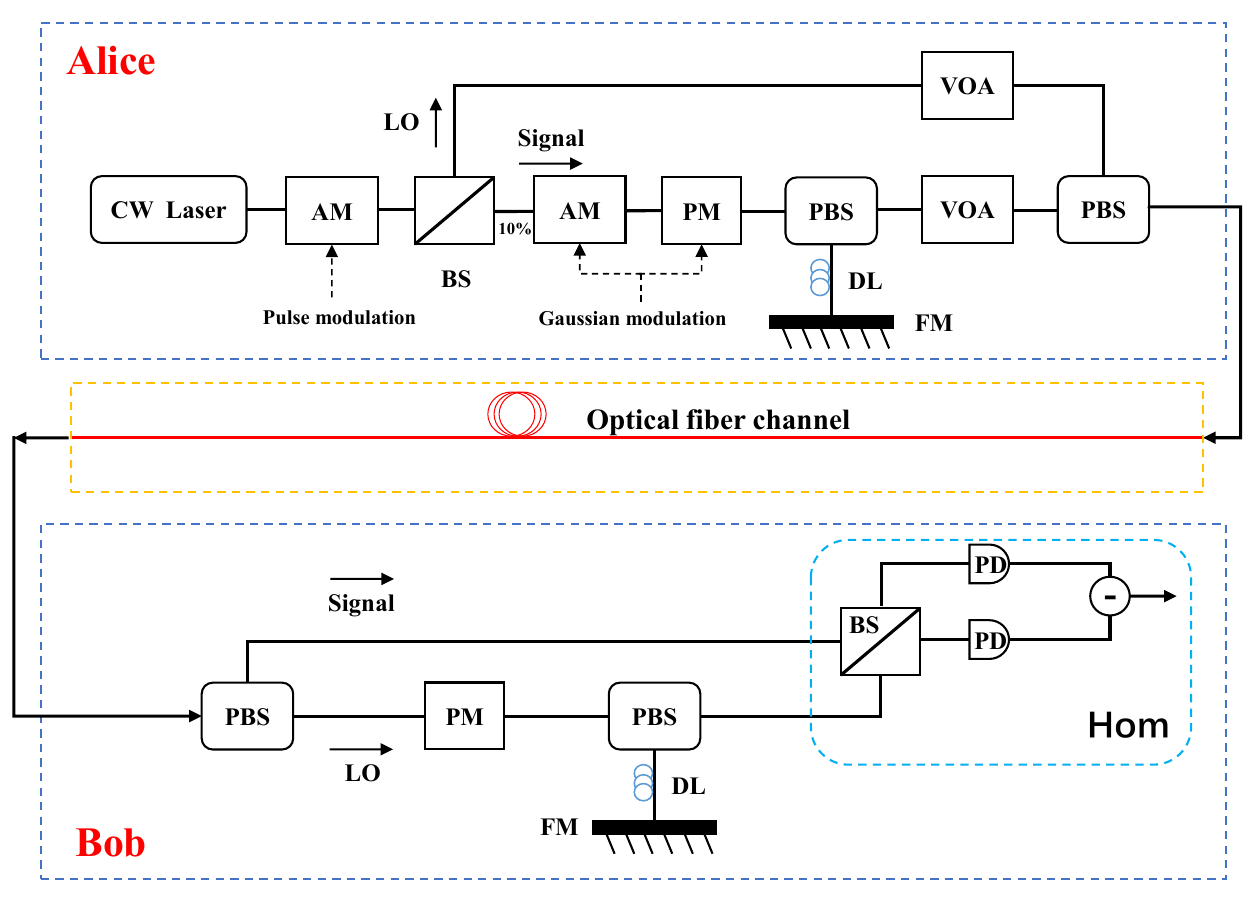}}
\caption{Practical optical path of the GMCS CVQKD system. AM, amplitude modulator; BS, beam splitter; LO, local oscillator; PM, phase modulator; PBS, polarization beam splitter; DL, delay line; FM, Faraday mirror; VOA, variable optical attenuator; Hom, homodyne detector; PD, photodetector; }
\label{FIG1}
\end{figure}

 Fig.\ref{FIG1} describes the standard CVQKD system. First, Alice prepares the initial coherent state using a commercial laser. Then, Alice uses a beam splitter (BS) to divide the initial coherent state into two parts: the signal path and the reference path. In the signal path, the signal passes through an amplitude modulator (AM) and a phase modulator (PM) to generate the Gaussian-modulated coherent state $\lvert x_{A} + i p_{A} \rangle$. Here, $x_{A}$ and $p_{A}$ are two independent random variables, and they follow the Gaussian distribution $N(0, V_{A})$, where $V_{A}$ is the modulation variance. In the reference path, i.e., the LO, the light passes through an optical attenuator and is then multiplexed with the prepared Gaussian-modulated coherent state in time-domain. Finally, the two lights are transmitted to Bob through a channel with transmittance $T$ and excess noise $\varepsilon_{0}$. At Bob, the received multiplexed signal is split into two paths, and the quadratures $X$ (position) and $P$ (momentum) are measured using a homodyne detector.

\begin{figure}[!h]\center
\centering
\resizebox{8cm}{!}{
\includegraphics{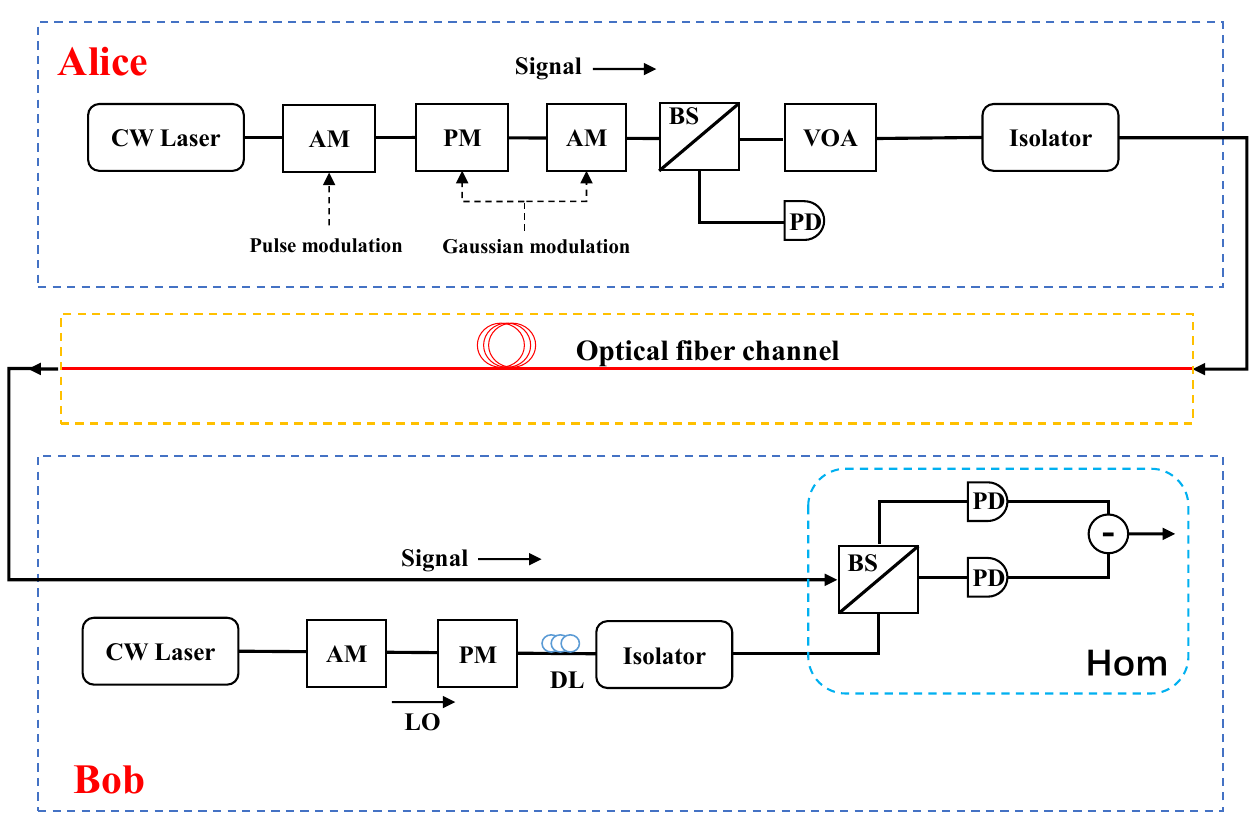}}
\caption{Optical schematic of an LLO-CVQKD system. Here, the LO is generated locally by Bob, and Alice does not need to transmit it.}
\label{FIG2}
\end{figure}

 For the LLO-CVQKD system, as shown in Fig.\ref{FIG2}, unlike the standard CVQKD implementation, since a separate laser is used at Bob to generate the LO, Alice does not need to send the LO. In an LLO-CVQKD system, because the LO and the signal are generated by different lasers, the system will generate more noise. For example, polarization-mismatch noise $\varepsilon_{PMN}$, phase noise $\varepsilon_{PN}$, and local-oscillator relative-intensity noise $\varepsilon_{RIN}$. We assume that these different noise sources are statistically independent, so the total excess noise can be expressed as the sum of all individual noise contributions. To simplify the analysis, the noise of the LLO-CVQKD system can be described as follows
 \begin{equation}\label{eq1}
    \begin{split}
    \varepsilon_{tot} = \varepsilon_{0} +\varepsilon_{PMN}+\varepsilon_{PN}+\varepsilon_{RIN}+ ...
    \end{split}
\end{equation}

Moreover, at Bob, a practical homodyne detector with detection efficiency $\eta$ and electronic noise variance $v_{\text{el}}$ is used for detection. Thus, we obtain the total noise referred to the channel input as $\chi_{tot} = \chi_{line}+ \chi_{hom} /T$, where \( \chi_{line} = 1/T - 1 + \varepsilon_{tot} \), and \( \chi_{hom} = [(1 - \eta) + v_{el}]/ \eta \).

\subsection{Impact of pulse width mismatch in a practical LLO-CVQKD system}
In above part, we introduce the LLO-CVQKD system. And in this section, we will describe the security issues that arise when there is a mismatch in the pulse width between the LO and the signal during detection on Bob.

In CVQKD systems, the optical fields of the signal and the LO are typically assumed to follow a Gaussian pulse form. This assumption not only simplifies theoretical analysis but also provides a good approximation of the characteristics of the light sources in actual experiments. Therefore, we assume that both the signal and the LO can be described by a single Gaussian temporal mode. Specifically, the temporal modes of the signal and the local oscillator are expressed as follows \cite{r38}

\begin{equation}\label{eq2}
    \begin{split}
    u_s(t) = \frac{1}{\sqrt{\tau_s} \pi^{1/4}} e^{-\frac{t^2}{2 \tau_s^2}},
    u_{LO}(t) = \frac{1}{\sqrt{\tau_{LO}} \pi^{1/4}} e^{-\frac{t^2}{2 \tau_{LO}^2}},
    \end{split}
\end{equation}
where $\tau_s$ and $\tau_{LO}$ are the pulse widths of the signal and the LO respectively. Moreover, $u_s(t)$ and $u_{LO}(t)$ each satisfy the normalization condition, that is

\begin{equation}\label{eq3}
    \begin{split}
    \int_{-\infty}^{\infty} |u_s(t)|^2 \, dt = 1,  \int_{-\infty}^{\infty} |u_{LO}(t)|^2 \, dt = 1.
    \end{split}
\end{equation}

\begin{figure}[!h]\center
\centering
\resizebox{8cm}{!}{
\includegraphics{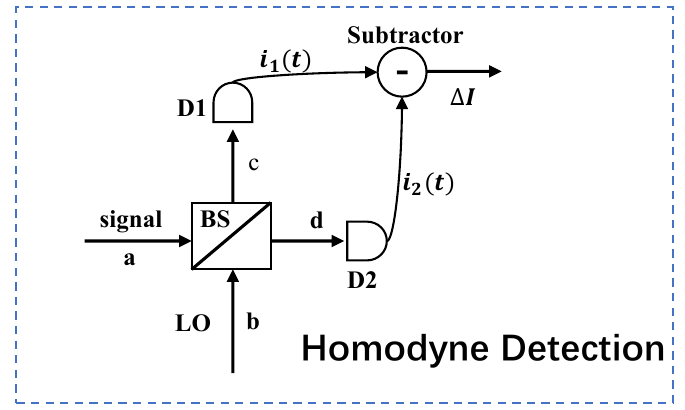}}
\caption{The structure of the homodyne detection. D1 and D2 are photodetectors.
 }
\label{FIG3}
\end{figure} 
We consider using the balanced homodyne detection. The structure of the balanced homodyne detection is shown in Fig.\ref{FIG3}. The signal and LO pass through a 50:50 beam splitter, and the two detectors output photocurrents $i_1(t)$ and $i_2(t)$. At points $c$ and $d$, we can obtain
\begin{equation}\label{eq4}
    \begin{split}
    \hat{c} = \frac{1}{\sqrt{2}} \left( \hat{a}_s + i \hat{a}_{LO} \right),
\hat{d} = \frac{1}{\sqrt{2}} \left( \hat{a}_s - i \hat{a}_{LO} \right),
    \end{split}
\end{equation}
When the pulse widths of the signal and the LO match, i.e., $\tau_s = \tau_{LO}$, we can calculate the current difference $\Delta I$
\begin{equation}\label{eq5}
    \begin{split}
    \Delta I &= i_1(t) - i_2(t) \\
             &=  i \hat{a}_s^\dagger \hat{a}_{LO} - i \hat{a}_{LO}^\dagger \hat{a}_s \\
             &=\hat{a}_s^\dagger \hat{a}_{LO} e^{\frac{\pi}{2}i} + \left( \hat{a}_{LO}e^{\frac{\pi}{2}i} \right)^\dagger  \hat{a}_s,
    \end{split}
\end{equation}
where $\hat{a}_{LO}^\dagger = |\alpha| e^{i\phi}$, and $\alpha$ is the amplitude of the local oscillator light.
Finally, we can get
\begin{equation}\label{eq6}
    \begin{split}
    \Delta I \propto |\alpha| ( X_s \cos \theta + P_s \sin \theta ).     
    \end{split}
\end{equation}

Thus, we can obtain $X_s$ and $P_s$ of the signal through detection. However, the above formula are only applicable to the perfect detection case where the pulse widths of the signal and the LO are matched. In a practical LLO-CVQKD system, the LO is prepared locally at Bob, rather than using a reference light sent from Alice. As a result, it is not emitted from the same CW laser as the signal. Furthermore, due to differences in the lasers or imperfections in the preparation process, this situation may lead to a mismatch in the pulse widths between the signal and the LO.

\begin{figure}[!h]\center
\centering
\resizebox{8cm}{!}{
\includegraphics{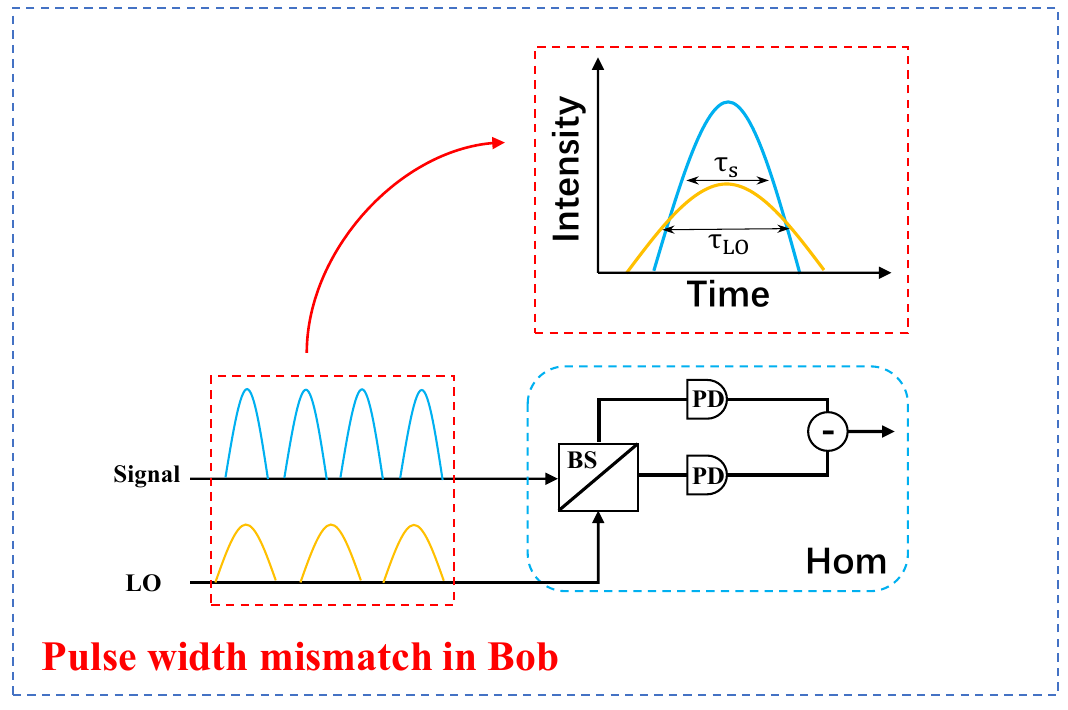}}
\caption{Actual situation of pulse width mismatch between the local oscillator and the signal.
}
\label{FIG4}
\end{figure}

We consider the case where the pulse widths of the LO and the signal are mismatched, i.e., $\tau_s \neq \tau_{LO}$. As shown in Fig.\ref{FIG4}, this mismatch affects Bob's measurement of the quantum state signal. According to Ref.\cite{r42} , we can derive
\begin{equation}\label{eq7}
    \begin{split}
    \hat{a}_s' = \gamma \hat{a}_s + \sqrt{1-\gamma^2} \hat{a}_\perp,
    \end{split}
\end{equation}
The formula for the differential current $\Delta I'$ then becomes
\begin{equation}\label{eq8}
    \begin{split}
    \Delta I' = &\gamma \left[ \hat{a}_s^\dagger \hat{a}_{LO} e^{\frac{\pi}{2}i} + \left( \hat{a}_{LO} e^{\frac{\pi}{2}i} \right)^\dagger \hat{a}_s \right] \\
    &+ \sqrt{1-\gamma^2} \left[ \hat{a}_\perp^\dagger \hat{a}_{LO} e^{\frac{\pi}{2}i} + \left( \hat{a}_{LO} e^{\frac{\pi}{2}i} \right)^\dagger \hat{a}_\perp \right],
    \end{split}
\end{equation}
where $\gamma  \in (0, 1]$ is the overlap coefficient, and we can calculate
\begin{equation}\label{eq9}
    \begin{split}
    \gamma &= \left| \int_{-\infty}^{+\infty} u_{LO}(t) u_s(t) \ dt \right|\\
    &=\sqrt{\frac{2 \tau_s \tau_{LO}}{\tau_s^2 + \tau_{LO}^2}}.
    \end{split}
\end{equation}
Thus, according to Eq.(\ref{eq4}) to (\ref{eq9}), we can obtain
\begin{equation}\label{eq10}
    \begin{split}
    x_B' = \gamma x_B + \sqrt{1-\gamma^2}x_\perp,
    \end{split}
\end{equation}

Since the mode perpendicular to $x$ is in the vacuum state, we can write the formula as
\begin{equation}\label{eq11}
    \begin{split}
    x_B' = \gamma x_B + \sqrt{1-\gamma^2}x_{vac}.
    \end{split}
\end{equation}
where $x_B$ is the value measured at Bob, and $x_B'$ is the value measured after considering the pulse width mismatch.
Similarly, we can also obtain
\begin{equation}\label{eq12}
    \begin{split}
    p_B' = \gamma p_B + \sqrt{1-\gamma^2}p_{vac}.
    \end{split}
\end{equation}

\begin{figure}[!h]\center
\centering
\resizebox{8cm}{!}{
\includegraphics{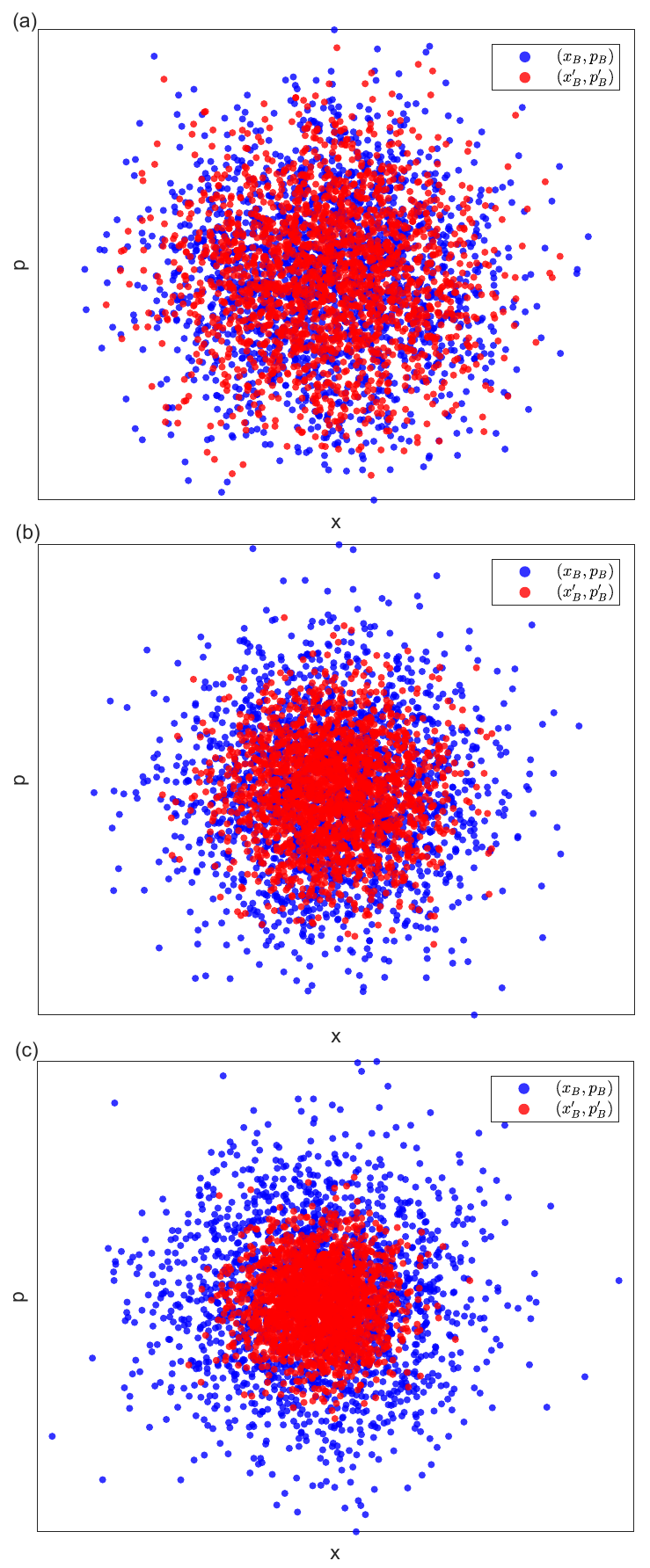}}
\caption{Corresponding changes in the measurement offsets of $x_B'$ ($p_B'$) relative to $x_B$ ($p_B$) with different $\gamma$ values.
(a) $\gamma=0.9$, (b) $\gamma=0.7$, (c) $\gamma=0.5$. 
 }
\label{FIG9}
\end{figure}

    From the above formulas, we can already observe that the pulse width mismatch between the signal and the LO has a important effect on Bob's measurements. The measurement values of $x_B$ and $p_B$ will vary with changes in $\gamma$, which may introduce a security loophole. Furthermore, as shown in Fig.\ref{FIG9}, we simulate the measurement offsets of $x_B$ and $p_B$ under the condition of pulse width mismatch. From these figures, it can be seen that as the $\gamma$ values decrease, the measured values of $x_B'$ and $p_B'$ deviate progressively further from their original true values. In other words, the more severe the pulse width mismatch, the less accurate the measured values of $x_B$ and $p_B$. And it results in reduced system performance and may lead to a loophole. In the next section, we perform parameter estimation and investigate how pulse width mismatch between the signal and LO affects the system’s security in a practical LLO-CVQKD system.

\section{PARAMETER ESTIMATION}\label{sec3}
In this section, we will conduct the task of parameter estimation and security analysis. Through parameter estimation, we can directly observe the problems that arise with pulse width mismatch.
After the balanced homodyne detection, Alice and Bob will share a group of correlated Gaussian data $X=\left\lbrace(x_{A_i},x_{B_i})|i=1,2,\ldots,N\right\rbrace$ or $P=\left\lbrace(p_{A_i},p_{B_i})|i=1,2,\ldots,N\right\rbrace$, where $N$ is the total number of the received pulses. In an LLO-CVQKD system, the quantum channel is assumed to be a linear model \cite{r39,r40,r41}, which can be represented as
\begin{equation}\label{eq13}
    x_{B}=tx_{A}+z,
\end{equation}
where $t=\sqrt{\eta T}$ and vector $z$ satisfies a centered Gaussian distribution with variance $\sigma^2=\eta T\xi+N_0+V_{el}$. Here, $\xi=\varepsilon_{tot} N_0$, $V_{el}=v_{el}N_0$, and $N_0$ is the variance of the shot noise. The parameter $T$ denotes the transmittance
of the quantum channel and $\eta$ is the efficiency of the
homodyne detectors.
According to eq. (\ref{eq13}), we can get the following expression
\begin{equation}\label{eq14}
    \begin{split}
        V_{A}=Var&(x_A)=\langle {x_A^2}\rangle,\cr
        V_{B}=Var(x_B)=\langle {x_B^2}\rangle&=\eta TV_A+\eta T\xi+N_0+V_{el},\cr
        Cov(x_A,x_B)&=\langle x_Ax_B\rangle=\sqrt{\eta T}V_A.
    \end{split}
\end{equation}
And due to symmetry, the relation between $p_A$ and $p_B$ has the same form as the above expression. When the pulse widths of the signal and the LO are mismatched, according to eq.(\ref{eq10}) and eq.(\ref{eq13}), we can obtain
\begin{equation}\label{eq15}
    \begin{split}
        &x_B'=\gamma tx_A + z',\cr
        V_B' = Var(x_B')&=\eta T \gamma^2 V_A + \eta T \gamma^2 \xi + \gamma^2V_{el} + N_0,\cr
        Cov(&x_A, x_B') = \gamma \sqrt{\eta T} V_{A}, \cr
    \end{split}
\end{equation}
And the variance of $z'$ is $\eta T \gamma^2 \xi + \gamma^2 V_{el} + N_0$.

In a practical system, Eve has no ability to alter Bob’s measurement of the signal. In other words, the parameters $\eta$ and $V_{el}$ remain unchanged.

If the signal is ideal and there is no pulse width mismatch between the signal and the local oscillator, then Alice and Bob will use the following expression for parameter estimation
\begin{equation}\label{eq16}
    \begin{split}
        T&=\frac{{Cov(x_A,x_B)}^2}{\eta V_A^2},\cr
        \xi&=\frac{V_B-N_0-V_{el}}{\eta T}-V_A.
    \end{split}
\end{equation}
However, when pulse width mismatch occurs, if Alice and Bob are unaware of this situation and still use eq.(\ref{eq16}) for estimation, then the result will be
\begin{equation}\label{eq17}
    \begin{split}
        T'&=\frac{{Cov(x_A,x_B^{\prime})}^2}{\eta V_A^2},\cr
        \xi'&=\frac{V_B^{\prime}-N_0-V_{el}}{\eta T'}-V_A.
    \end{split}
\end{equation}
After further manipulation of the above equations, we obtain
\begin{equation}\label{eq18}
    \begin{split}
        T'=\gamma^2T,
        \varepsilon_{tot}'= \varepsilon_{tot} - \frac{(1-\gamma^2)v_{el}}{\eta T \gamma^2}. (0<\gamma\leqslant  1) 
    \end{split}
\end{equation}

\begin{figure}[!h]\center
\centering
\resizebox{8cm}{!}{
\includegraphics{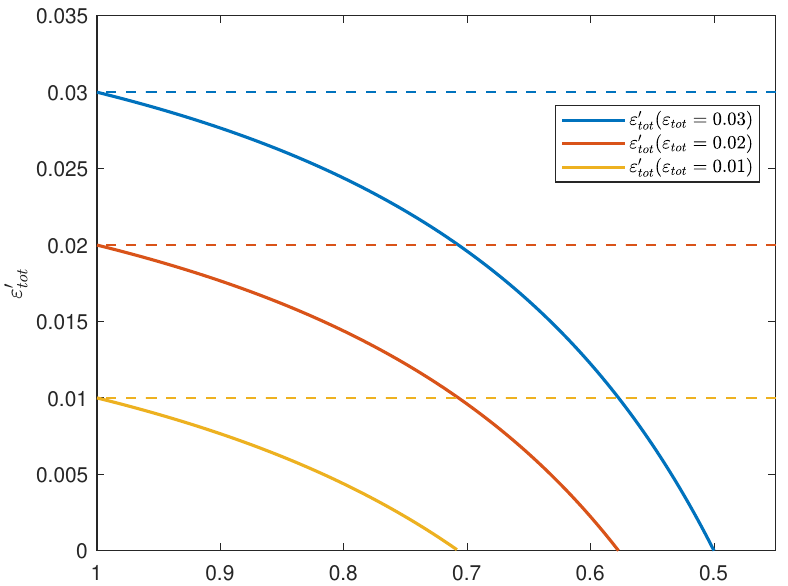}}
\caption{The variation of noise $\varepsilon_{tot}$ with the value of $\gamma$. The dashed lines in the corresponding colors represent the values when $\gamma$ is always equal to 1.
}
\label{FIG5}
\end{figure}

By observation, it can be seen that both the transmittance of the quantum channel and the total excess noise of the system are incorrectly estimated.  As shown in Fig.\ref{FIG5}, with the variation of $\gamma$, the estimated value of the noise $\varepsilon_{tot}'$ becomes lower than the ideal value, and this gives rise to security issues. Next, we use the classical partial intercept-resend (PIR) attack as an example to analyze the security of a practical LLO-CVQKD system under the effects of pulse width mismatch.

When Eve performs a PIR attack, the probability distribution of Bob’s detection results can be expressed as a weighted sum of two Gaussian distributions \cite{r43,r44}. The first part corresponds to the distribution of the data resent by Eve, with weight $u$; the second part corresponds to the distribution of the data transmitted by Alice, with weight $1-u$. Furthermore, the noise introduced by the PIR attack is $2uN_0$. Therefore, when Eve executes a PIR attack, the theoretical upper bound of the ideal estimation value of the excess noise of the quantum channel should be
\begin{equation}\label{eq19}
    \begin{split}
        \varepsilon_{tot}^{PIR} = \varepsilon_{tot} + 2u,
    \end{split}
\end{equation}
Here, we use $u = 0.2$ to analyze the PIR attack in a general situation. Correspondingly, the estimated excess
noise becomes $\varepsilon_{tot}  + 0.4$. In this case, when the pulse widths of the local oscillator and the signal are mismatched, the measured noise can be expressed as
\begin{equation}\label{eq20}
    \begin{split}
        {\varepsilon_{tot}^{PIR}}' = \varepsilon_{tot} + 0.4 - \frac{(1-\gamma^2)v_{el}}{\eta T \gamma^2}.
    \end{split}
\end{equation}

In a practical LLO-CVQKD system, suppose the total excess noise is 0.1. Then, when Eve carries out a PIR attack and alters the pulse width of the signal, the estimated excess noise of the quantum channel should be ${\varepsilon_{tot}^{PIR}}' =0.5 - \frac{(1-\gamma^2)v_{el}}{\eta T \gamma^2}$. Under ideal conditions, when $\gamma = 1$, then there is
${\varepsilon_{tot}^{PIR}}' =\varepsilon_{tot} = 0.5.$
That is, the estimated excess noise equals the actual total excess noise, since there is no pulse width mismatch between the signal and the local oscillator. However, when $\eta$, $T$, and $V_{el}$ are fixed, the noise will be underestimated depending on the value of $\gamma$. For example, when $\gamma = 0.8452$ and $\tfrac{V_{el}}{\eta T} = 1$, excess noise will be 0.1 closely, which is equal to the actual excess noise. Moreover, Eve can even further manipulate the value of $\gamma$, making the noise value smaller. In particular, when $u = 1$, that is, when Eve performs a full intercept-resend attack, he can still alter the value of $\gamma$ to conceal his actions. 

Based on the above analysis, when Eve intercepts the signal in the channel, she can use certain tools to change the pulse width of the signal, or she can regenerate a new signal as a replacement with a different pulse width. By altering the pulse width of the signal, Eve can exploit the underestimation of the total excess noise induced by pulse width mismatch to hide her intercept–resend attack. Such an attack is practicable which seriously destroys the security of the practical system. Next, we will further examine it from the perspective of the secret key rate.

\section{Secret key rate under the pulse width mismatch}\label{sec4}
After parameter estimation, Alice and Bob will utilize $n$ received pulses to establish the secret key of the GMCS CVQKD system. In the case of collective attacks, the theoretical secret key rate of the system considering the reverse reconciliation and finite-size effect can be expressed as \cite{r29, r45}
\begin{equation}\label{eq21}
    K=\frac{n}{N}[\beta I_{A\!B}-S_{B\!E}^{\epsilon_{P\!E}}-\Delta(n)],
\end{equation}
where $n=N-m$ and $\beta$ is the reverse reconciliation efficiency. The mutual information $I_{AB}$ between Alice and Bob can be represented as
\begin{equation}\label{eq22}
    I_{A\!B}=\frac{1}{2}\log_2{\frac{V_B}{V_{B|A}}}=\frac{1}{2}\log_2{\frac{V+\chi_{tot}}{1+\chi_{tot}}},
\end{equation}
where $V=V_A+1$, $\chi_{tot}=\chi_{line}+\chi_{hom}/T$ is the total noise referred to the channel input, $\chi_{line}=1/T-1+\varepsilon_{tot}$ is the total channel-added noise referred to the channel input, and $\chi_{hom}=[(1-\eta)+v_{el}]/\eta$ is the detection-added noise referred to Bob’s input. $S_{B\!E}^{\epsilon_{P\!E}}$ is the maximum value of the Holevo information compatible with the statistics except with probability $\epsilon_{P\!E}$. In particular, the covariance matrix between Alice and Bob is related to $S_{B\!E}^{\epsilon_{P\!E}}$, which can be calculated as
\begin{equation}\label{eq23}
    \begin{split}
        \Gamma_{A\!B}&=
        \left[\begin{array}{ccc}
            \Gamma_A & \sigma_{A\!B}^T\cr
            \sigma_{A\!B} &\Gamma_{B}
        \end{array}
        \right]\cr
        &=
        \left[\begin{array}{ccc}
            VI_2 & \sqrt{T_{min}(V^2-1)}\sigma_z\cr
            \sqrt{T_{min}(V^2-1)}\sigma_z & [T_{min}(V+\chi_{line,max})]I_2
        \end{array}
        \right],
    \end{split}
\end{equation}
where $I_2=diag[1,1]$, $\sigma_z=diag[1,-1]$, $\chi_{line,max}=1/{T_{min}}-1+\varepsilon_{max}$, and $T_{min}$ and $\varepsilon_{max}$ correspond to the lower bound of $T$ and the upper bound of $\varepsilon$, respectively.

Based on the analysis in Sec.\ref{sec3}, the quantum channel involved in an LLO-CVQKD system is assumed to be a linear model and the parameters $T$ and $\xi$ are estimated by using $m$ pairs data from the $X$ or $P$. When $m$ is large enough (e.g., $m>10^6$), $T_{min}$ and $\varepsilon_{max}$ can be calculated as \cite{r45}
\begin{equation}\label{eq24}
    \begin{split}
        T_{min}&=\frac{(\hat{t}-\Delta t)^2}{\eta},\cr
        \varepsilon_{max}&=\frac{{\hat{\sigma}}^2+\Delta {\sigma}^2-N_0-v_{el}N_0}{{\hat{t}}^2N_0}.
    \end{split}
\end{equation}
For a linear model, the maximum-likelihood estimators $\hat{t}$ and ${\hat{\sigma}}^2$ can be expressed as
\begin{equation}\label{eq25}
    \hat{t}=\frac{\sum_{i=1}^{m}{x_A}_i{x_B}_i}{\sum_{i=1}^{m}{{x^2_A}_i}},\qquad{\hat{\sigma}}^2=\frac{1}{m}\sum_{i=1}^{m}({x_B}_i-\hat{t}{x_A}_i)^2.
\end{equation}
Additionally, $\Delta t$ and $\Delta{\sigma}^2$ can be calculated as
\begin{equation}\label{eq26}
    \Delta t=z_{\epsilon_{P\!E}/2}\sqrt{\frac{{\hat{\sigma}}^2}{mV_{x_A}}},\qquad\Delta{\sigma}^2=z_{\epsilon_{P\!E}/2}\frac{{\hat{\sigma}}^2\sqrt{2}}{\sqrt{m}}.
\end{equation}
where the coefficient $z_{\epsilon_{P\!E}/2}$ satisfies the following relation
    $1-\frac{1}{2}er\!f(z_{\epsilon_{P\!E}/2}/\sqrt{2})={\epsilon_{P\!E}}/2$, and $er\!f(\cdot)$ is the error function which can be expressed as $er\!f (x)=2\pi^{-\frac{1}{2}}\int_{0}^{x}e^{-t^2}dt$.

Then $S_{B\!E}^{\epsilon_{P\!E}}$ can be calculated as
\begin{equation}\label{eq27}
    S_{B\!E}^{\epsilon_{P\!E}}=\sum_{i=1}^{2}G\Big(\frac{\lambda_i-1}{2}\Big)-\sum_{i=3}^{5}G\Big(\frac{\lambda_i-1}{2}\Big),
\end{equation}
where $G(x)=(x+1)\log_2(x+1)-x\log_2x$ and $\lambda_i$ are the symplectic eigenvalues of the covariance matrix between Alice and Bob, which can be expressed as
\begin{equation}\label{eq28}
    \begin{split}
        \lambda_{1,2}^{2}&=\frac{1}{2}(A\pm\sqrt{A^2-4B}),\cr
        \lambda_{3,4}^{2}&=\frac{1}{2}(C\pm\sqrt{C^2-4D}),\cr
        \lambda_5&=1.
    \end{split}
\end{equation}
Here
\begin{equation}
    \label{eq29}
    \begin{split}
        A=&det\Gamma_{\!A}\!+\!det\Gamma_{\!B}\!+\!2det\sigma_{\!A\!B}\cr
        \quad=&V^2(1-2T_{min})+2T_{min}+{T_{min}}^2(V+\chi_{line,max})^2,\cr
        B=&det\Gamma_{\!A\!B}={T_{min}}^2(V\chi_{line,max}+1)^2,\cr
        C=&\frac{A\chi_{hom}+V\sqrt{B}+T_{min}(V+\chi_{line,max})}{T_{min}(V+\chi_{line,max}+\chi_{hom}/T_{min})},\cr
        D=&\sqrt{B}\frac{V+\sqrt{B}\chi_{hom}}{T_{min}(V+\chi_{line,max}+\chi_{hom}/T_{min})}.
    \end{split}
\end{equation}

Furthermore, in a practical CVQKD system, $\Delta(n)$ is related to the security of the privacy amplification, which can be written as
\begin{equation}
    \label{eq30}
\Delta(n)=7\sqrt{\frac{\log_2(1/\bar{\epsilon})}{n}}+\frac{2}{n}\log_2{\frac{1}{\epsilon_{P\!A}}},
\end{equation}
where $\bar{\epsilon}$ and $\epsilon_{P\!A}$ represent the smoothing parameter and the failure probability of privacy amplification, respectively. And because the value of $\Delta(n)$ mainly depends on $n$, the values of $\bar{\epsilon}$ and $\epsilon_{P\!A}$ are usually set to be equal to the value of $\epsilon_{P\!E}$.

\begin{figure}[!h]\center
\centering
\resizebox{8cm}{!}{
\includegraphics{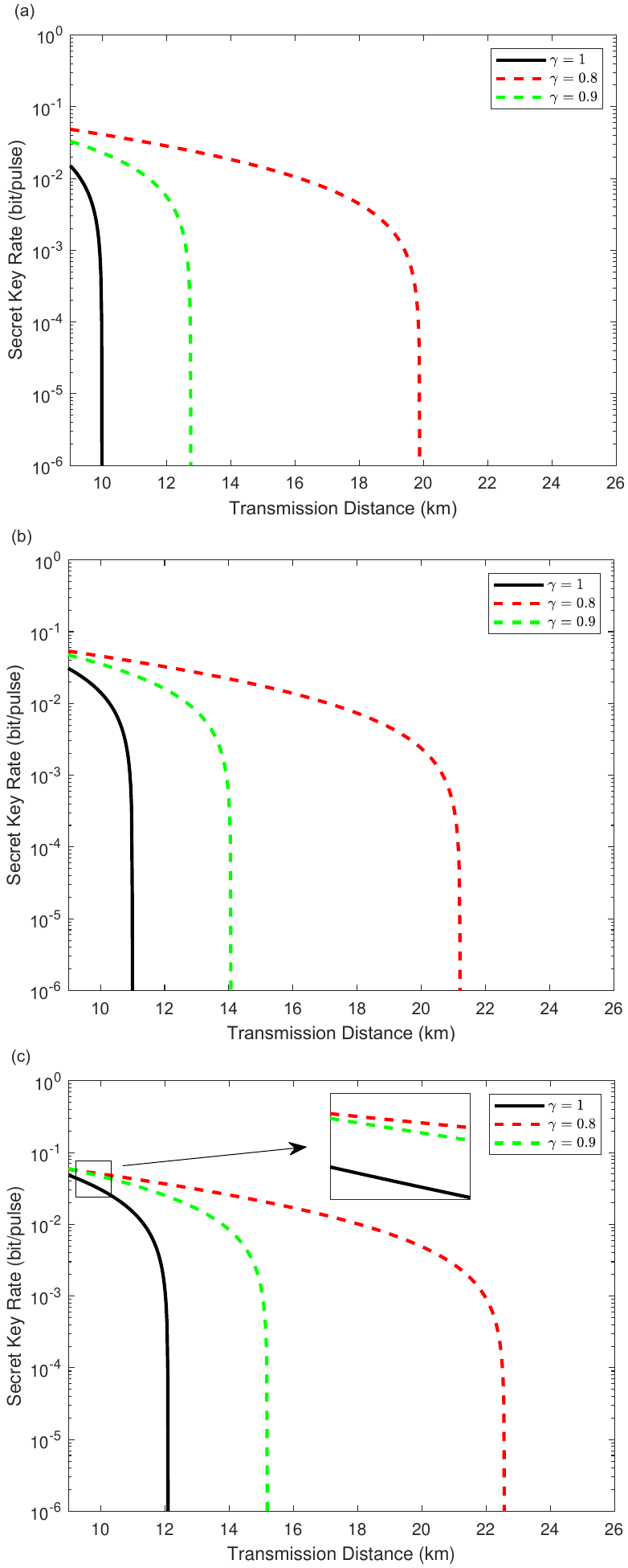}}
\caption{Secret key rate versus transmission distance when the pulse widths of the signal and LO are mismatched under different $\gamma$. The fiber loss is 0.2 dB/km. (a) $\varepsilon_{tot}=0.04$, (b) $\varepsilon_{tot}=0.03$, (c) $\varepsilon_{tot}=0.02$.}
\label{FIG6}
\end{figure}

\begin{figure}[!h]\center
\centering
\resizebox{8cm}{!}{
\includegraphics{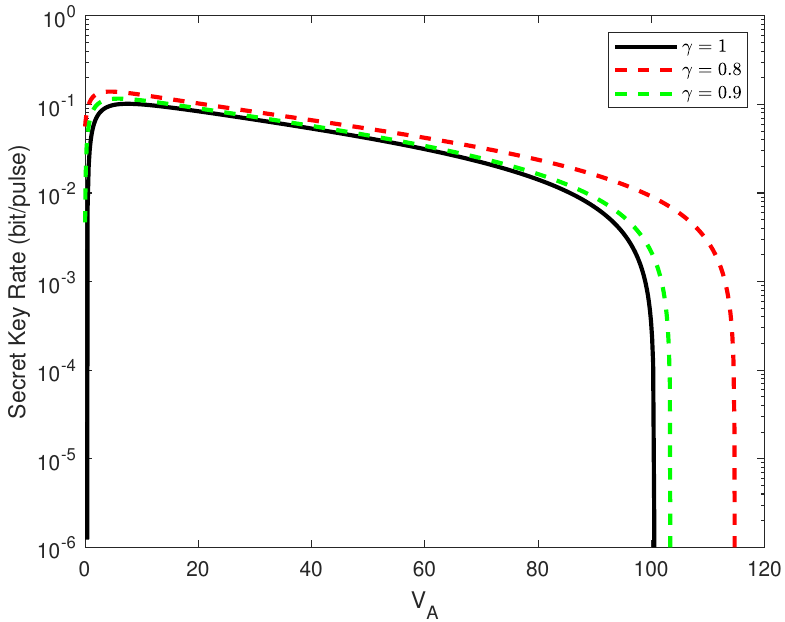}}
\caption{Secret key rate versus $V_A$ when the pulse widths of the signal and LO are mismatched under different $\gamma$. The transmission distance $L$ is 7km and $\varepsilon_{tot}=0.04$.}
\label{FIG7}
\end{figure}

According to the above description and formula, the secret key rate can be expressed as $K = K(V_A, T, \varepsilon_{tot}, \eta, v_{el})$. In the Sec.\ref{sec3}, we analyze the changes in parameter estimation when the pulse widths of the LO and the signal are mismatched. We find that among the parameters involved in the secret key rate calculation, $T$ and $\varepsilon_{tot}$ have changed, according to eq.(\ref{eq17}) and eq.(\ref{eq18}). In this way, the secret key rate calculated through parameter estimation becomes
$K' = K(V_A, T', \varepsilon_{tot}', \eta, v_{el})$. To further describe the difference between the measured secret key rate $K'$ and the actual $K$, we carry out simulations under the condition of pulse width mismatch. 
As shown in Fig.\ref{FIG6}, we simulate the secret key rate with distance at different mismatch levels, and the fixed parameters for the simulation are set as
$V_A = 40, \ \eta = 0.9, \ v_{el} = 0.1, \ T = 10^{-\alpha L/10}, \ \alpha = 0.2 \,\text{dB/km}, \ \beta = 0.8, \ N = 10^9, \ n = 0.5 \times N, \ \bar{\epsilon} = \epsilon_{PA} = 10^{-10}$. Moreover, in Fig.\ref{FIG7}, we simulate the key rate varying with $V_A$, and distance $L=7$;

From Fig.\ref{FIG6}, we can observe that when $\gamma$ is less than 1, the measured values of the secret key rate and the maximum transmission distance are higher than the true value ($\gamma=1$). Moreover, as $\gamma$ decreases further, this discrepancy in the measured values becomes more higher. And in Fig.\ref{FIG7}, we can observe similar results that the higher the level of mismatch, the more overestimated the key rate becomes. 

\begin{figure}[!h]\center
\centering
\resizebox{8cm}{!}{
\includegraphics{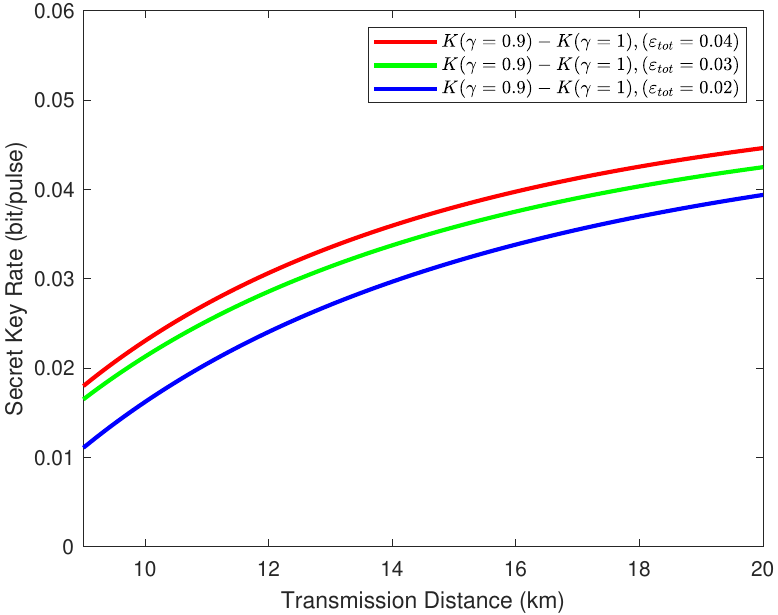}}
\caption{Difference between the estimated secret key rate $K(\gamma=0.9)$ and the practical secret key rate $K(\gamma=1)$.}
\label{FIG11}
\end{figure}

\begin{figure}[!h]\center
\centering
\resizebox{8cm}{!}{
\includegraphics{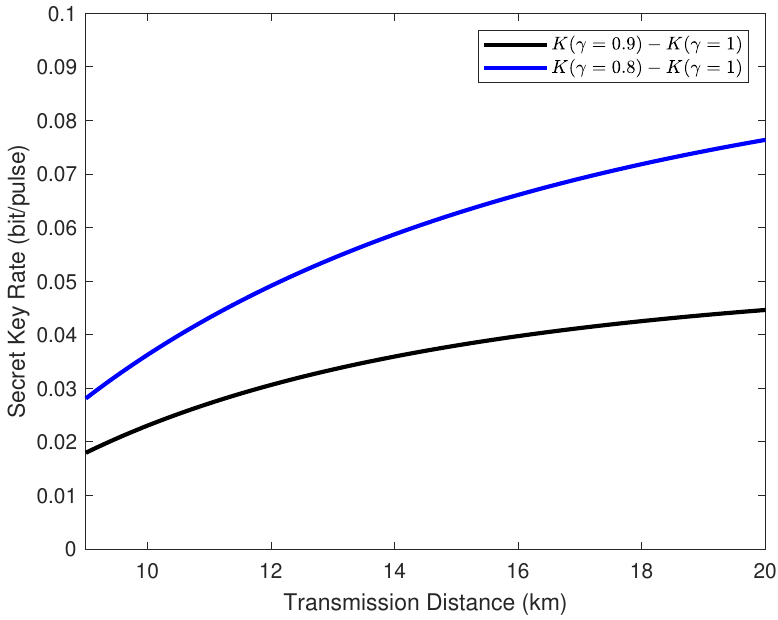}}
\caption{Difference between the estimated secret key rate $K$ with different $\gamma$ and the practical secret key rate $K(\gamma=1)$. $\varepsilon_{tot}=0.04$.}
\label{FIG12}
\end{figure}

Furthermore, to further analyze the discrepancy between the measured and actual secret key rates under different levels of total excess noise, in Fig.\ref{FIG11}, we simulate the difference between the secret key rate with pulse width mismatch and the actual value under different total excess noise conditions. And to clearly show the difference between the calculated secret key rate under different mismatch levels and the true key rate, as shown in Fig.\ref{FIG12}, we also simulate the difference between the key rate with different $\gamma$ values and the true key rate , while keeping the total excess noise the same. We find that, compared to lower total excess noise, the same level of pulse width mismatch will result in a more significant overestimation of the secret key rate in the case of higher excess noise. In other word, the larger the total excess noise, the greater the impact of pulse width mismatch on the security of the system. Besides, form Fig.\ref{FIG12}, a higher level of pulse width mismatch provides more opportunity for Eve to launch attacks. Moreover, as the degree of pulse width mismatch increases, the overestimation of the secret key rate also worsens with the increase in transmission distance.

These results indicate that pulse width mismatch can lead to an overestimation of the secret key rate, allowing Eve to conceal her attack by altering the pulse width of the intercepted signal. In the next section, we introduce a countermeasure to address this security issue.

\section{Countermeasure}\label{sec5}
In the above sections, we describe and analyze the security issues that arise when the LO and the signal have a mismatch in pulse width, and further examine the problem from the perspective of the secret key rate. This mismatch introduces a security loophole to the system. Therefore, it is necessary to seek methods to close the loophole. In this section, we introduce a defense strategy to counter such a situation.

\begin{figure}[!h]\center
\centering
\resizebox{8cm}{!}{
\includegraphics{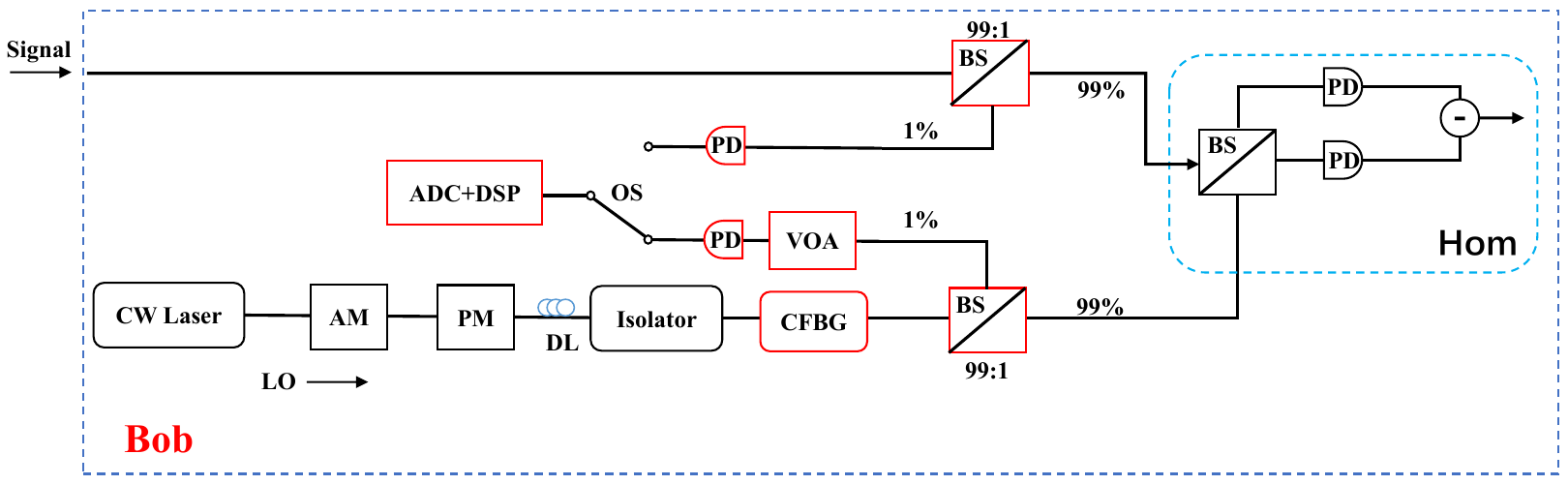}}
\caption{A scheme for monitoring and correcting the pulse widths of the LO and signal in an LLO-CVQKD system. CFBG, chirped fiber bragg grating; ADC, analog-to-digital converter; DSP, digital signal processor; OC, optical swith. 
}
\label{FIG8}
\end{figure}

In a standard LLO-CVQKD system, there are generally no devices for real-time monitoring of pulse width or related information.
However, variations in the pulse width of the LO, or of the signal, directly affect the results of parameter estimation. Therefore, as illustrated in the Fig.\ref{FIG8}, we first design a monitoring scheme. In this measure, before the signal is transmitted to Bob through the channel, we extract 1\% of the LO for analysis. Since the LO intensity may be too strong, the extracted LO is first reduced in intensity with an optical attenuator. Next, we use an analog-to-digital converter (ADC) to convert the light into a digital form, and then use a digital signal processor (DSP) to determine the pulse width of the LO. When the signal enters Bob from the channel, we extract 1\% of it using the same method. Since the signal intensity is not as strong as LO, we don't need an optical attenuator. Then, by configuring the optical switch (OC), we use the ADC and DSP to measure the pulse width of the signal. In this way, we obtain the pulse widths of both the signal and the LO. Next, their data can be compared in the DSP. If the pulse widths are the same, no additional processing is required; if they differ, the shape of the LO needs to be corrected.

To further optimize the matching between the signal and the LO, and to improve the measurement accuracy of the system, we introduce a chirped fiber bragg grating (CFBG) \cite{r46}. By properly configuring the CFBG, we can precisely control the temporal and spectral characteristics of the optical pulses, and achieve higher-precision monitoring and defense strategies for the practical LLO-CVQKD system. A CFBG is an optical device made by periodically modulating the refractive index within the fiber, which can selectively reflect light within a specific wavelength range. By adjusting the temperature, voltage, or optical control, its reflection spectrum or group delay can be altered, enabling the stretching or compression of optical pulses. If a pulse width mismatch occurs, we can appropriately compress or stretch the LO pulse using a CFBG, thereby restoring the pulse widths to match. This way, it can be ensured that the local oscillator light and the signal have matching pulse widths before entering the measurement, i.e., $\tau_{LO} = \tau_{s}$, thereby eliminating the loophole.

\section{CONCLUSION}\label{sec6}

In this paper, we investigate the impact of pulse width mismatch between the signal and the LO on LLO-CVQKD systems. 
We establish a model for pulse width mismatch in LLO-CVQKD systems and then further analyze how this mismatch affects parameter estimation and how Eve can exploit it to perform an attack. In addition, we also simulate how this mismatch affect Bob’s measurements and the secret key rate. Based on the analysis, We find that when such a mismatch occurs, the key parameter $T$ and $\varepsilon_{tot}$ involved in the calculation of the secret key rate are incorrectly estimated. Moreover, as the severity of the pulse width mismatch increases, the noise is increasingly underestimated. This leads to an overestimation of the overall system’s secret key rate, creating potential security loopholes and providing Eve with opportunities to attack.

Moreover, we propose a solution by designing a scheme for monitoring and correcting the pulse width. The countermeasure we proposed extracts small portions of light from both the signal and LO paths at Bob. Through the joint operation of the added analog-to-digital converter and digital signal processor, the scheme analyzes and compares the pulse widths of the two signals. Moreover, we set up optical switches to simplify the system structure to minimize repeated use of devices. After comparing the pulse widths, if a mismatch is detected, we add a new device, a CFBG, in the LO path. By controlling the delay of the reflected light, the CFBG broadens or compresses the pulses, ultimately realigning the pulse widths of the LO and the signal. By readjusting the pulse width of the LO, we ensure that the pulse widths of the signal and the LO are identical before entering the detection, i.e., $\tau_{LO} = \tau_s$, successfully eliminating this security loophole. In addition, the scheme is expected to provide a more stable and efficient solution for large-scale continuous-variable quantum key distribution systems, advancing the practical development of quantum communication.

\section*{Acknowledgments}
This work was supported by the Joint Funds of the National Natural Science Foundation of China under Grant No. U22B2025, Key Research and Development Program of Shaanxi under Grant No. 2024GX-YBXM-077, the Stability Program of National Key Laboratory of Security Communication under Grant No. WD202406 and the Fundamental Research Funds for the Central Universities under Grant No. D5000210764.

\end{document}